\documentclass[twocolumn,12pt,cites,graphicx]{article}
\usepackage{epsfig,amsmath,amssymb}
\usepackage{mathrsfs}
\usepackage{bm}% bold math
\newcommand{\D}{\mathscr{D}}

\title{Interface dynamics in shear-banding flow of giant micelles}
\author{S. Lerouge$^{a}$\thanks{sandra.lerouge@univ-paris-diderot.fr}, M.A. Fardin$^{a}$, M. Argentina$^{b}$, G. Gr\'egoire$^{a}$, O. Cardoso$^{a}$\\\\ $^{a}$Laboratoire Mati\`ere et Syst\`emes Complexes, Universit\'e Paris-Diderot,\\ UMR 7057 CNRS, 10 rue Alice Domon et L\'eonie Duquet, 75205 Paris C\'edex 13\\\\$^{b}$Institut Non Lin\'eaire, Universit\'e de Nice Sophia Antipolis, UMR 6618 CNRS,\\ 1361 Route des Lucioles, 06560 Valbonne, France}

\date{Received XXXXth Month, 200X\\Accepted XXXXth Month,200X\\DOI: 10.1039/}

\begin{document}

\maketitle
\renewcommand{\thefootnote}{\fnsymbol{footnote}}

\noindent We report on a non trivial dynamics of the interface between shear bands following a start-up of flow in a semi-dilute wormlike micellar system investigated using a combination of mechanical and optical measurements. During the building of the banding structure, we observed the stages of formation, migration of the interface between bands and finally the destabilization of this interface along the vorticity axis. The mechanical signature of these processes has been indentified in the time series of the shear stress. The interface instability occurs all along the stress plateau, the asymptotic wavelength of the patterns increasing with the control parameter typically from a fraction of the gap width to about four times the gap width. Three main regimes of dynamics are highlighted : a spatially stable oscillating mode approximately at the middle of the coexistence region flanked by two ranges where the dynamics appears more exotic with propagative and chaotic events respectively at low and high shear rates. The distribution of  small particles seeded in the solution strongly suggests that the flow is three-dimensional. Finally, we demonstrate that the shear-banding scenario described in this paper is not specific to our system.

\section{Introduction}\label{intro}
Many complex fluids often show original non linear responses when submitted to hydrodynamic forces even of low intensity. These non linear behaviours resulting from the coupling between the internal structure of the fluid and the flow  are usually associated with a new mesoscopic organization of the system. In turn, the modification of the supramolecular architecture of the fluid can affect the flow itself and, for example, generate shear localization effects generally characterized by a splitting of the system into two macroscopic layers bearing different shear rates and stacked along the velocity gradient direction. This transition towards a heterogeneous flow has been reported in complex fluids of various microstructure such as lyotropic micellar and lamellar phases \cite{Ber3,Sal2}, triblock copolymers solutions \cite{Ber5,Scho1}, viral suspensions \cite{Let1}, thermotropic liquid crystal polymers \cite{Puj1}, electro-rheological fluids \cite{Vol1}, soft glassy materials \cite{Cou1}, granular materials \cite{Los1,Mue1}  or foams \cite{Deb1,Den1,Den2}. Among these systems, the shear banding flow of reversible giant micelles is particularly well documented \cite{Ber3}. The rheological signature of this type of flow has been observed for the first time in the pioneering work of Rehage et al \cite{Reh1}: the measured flow curve $\sigma=f(\dot\gamma)$ is composed of two stable branches respectively of high and low viscosities separated by a stress plateau  at $\sigma=\sigma_{p}$ extending between two critical shear rates $\dot\gamma_{l}$ and $\dot\gamma_{h}$. When the imposed shear rate $\dot\gamma$ is lower than $\dot\gamma_{l}$, the state of the system is described by the high viscosity branch which is generally shear-thinning : the micellar threads are slightly orientated and the flow is homogeneous. For macroscopic shear rates above $\dot\gamma_{l}$, the flow becomes unstable and evolves towards a banded state where the viscous and fluid phases coexist at constant  stress $\sigma_{p}$. The modification of the control parameter is supposed to only affects the relative proportions $1-\alpha_{h}$ and $\alpha_{h}$ of each band according to the lever rule $\dot\gamma=(1-\alpha_{h})\dot\gamma_{l}+\alpha_{h}\dot\gamma_{h}$ that results from the continuity of the velocity at the interface. Above $\dot\gamma_{h}$, the system is entirely converted into the fluid phase : the induced structures are strongly aligned along the flow direction and the homogeneity of the flow is recovered. This scenario has been predicted by Cates and coworkers more than ten years ago  \cite{Cat1}. Recently, different velocimetry techniques have been developped to determine the velocity profiles in a plane velocity/velocity gradient resolved both in time and space \cite{Sal0,Man0,Lopez1}. In particular Salmon et al demonstrated using dynamic light scattering that the CPCl/NaSal system, extensively studied during the last decade, was following the classical shear-banding organization  described above \cite{Sal1}.\\
Besides, more complex banding pictures on slightly different systems also emerged from the combination of macroscopic rheology together with local measurements.  Several groups established the existence of temporal fluctuations of the local flow field using nuclear magnetic resonance (NMR) \cite{Holm1,Lopez1}, particle image velocimetry (PIV) \cite{Hu1} or high frequency ultrasonic velocimetry \cite{Bec1}. They also mentioned oscillations of the interface position as function of time, sometimes correlated with  wall slip effects.  Moreover, flow birefringence experiments which give information on the averaged orientation of the medium also show clear evidence of local ordering fluctuations in the banding structure \cite{Ler1,Ler4,Hu1}. In some cases, these complex spatio-temporal evolutions are associated with fluctuations in the shear stress time series \cite{Holm1,Lopez1,Band1,Gan1}.\\
Very recently, several theoretical studies based on the derivation of the modified Johnson-Segalman equation using perturbation analysis tempted to rationnalize these fluctuating behaviours. They mainly focused on the stability of the shear-banded state in planar flow with respect to small perturbations with wave vector in the plane made by the flow and the vorticity directions \cite{Fiel1, Wils1,Fiel2}. For sufficiently thin interfaces, the banding state is found to be linearly unstable, except for extremely low and high shear rates in the plateau region, with respect to perturbations with wave vector in the flow direction. The parameters driving the instability are jumps in normal stresses and shear rate accross the interface. The non linear analysis reveals a complex spatio-temporal dynamics of the interface with a transition from travelling to rippling waves depending on the ratio between the thickeness of the interface and  the length of the cell. The authors also mentioned the existence of a modulation  of the interface along the vorticity direction with a much slower growth rate but a non negligible contribution to the asymptotic state \cite{Fiel3}. Subsequently, these undulations generate velocity rolls stacked along the vorticity direction. \\
In a previous study, we reported on the  behaviour of the interface in the shear-banding flow of a sample made of cetyltrimethylammonium bromide and sodium nitrate in Couette geometry \cite{Ler5}. Using the scattering properties of each band, we showed that the interface between bands becomes unstable with the wave vector in the direction of the cylinders axis. \\
In the present paper, we study the interface dynamics of this same sample in a detailed way. We carefully examine the formation and destabilization of the interface together with the temporal evolution of the shear stress to identify the mechanical signature of these different processes. We also explore the stress plateau in order to properly characterize the spatio-temporal dynamics: three main regimes were identified, with propagative events at low shear rates, spatially stable undulations for intermediate controlled parameter and finally chaotic pattern at high shear rates. The asymptotic wavelength is found to increase with the mean shear rate whereas the variation of the amplitude of the interface profile is non-monotonic. The three-dimensional character of the flow field is  strongly suggested by the organization of small tracers embedded in our solution into stripes stacked along the vorticity direction. Finally, we show that this interface instability is not particular to our sample but is also present in the CPCl/NaSal system studied, among others, in references \cite{Sal1,Hu1}. Our results are discussed by the light of the recent litterature on the stability of shear banded flows \cite{Fiel3} and in the context of purely elastic instabilities \cite{Sha1,Groisman2}.
\section{\label{exp}Experimental details}
\subsection{\label{mat}Materials}
The micellar sample under investigation is made of a classical surfactant, the cetyltrimethylammonium bromide (CTAB) at 0.3M mixed with a mineral salt, the sodium nitrate (NaNO$_{3}$) at 0.405M, in distilled water. Both the surfactant and the salt are supplied by Acros Organics and are used without further purification. The temperature is fixed at T=28$^{\circ}$C. At the concentration chosen here (around 11\% wt), far from the isotropic/nematic transition at rest, the solution is semi-dilute and made of highly entangled wormlike micelles forming an elastic network. In the linear regime, this network behaves as a purely Maxwellian element  with a single relaxation time $\tau_{R}=0.23\pm 0.02s$ and a plateau modulus $G_{0}=238\pm 5 Pa$. Under simple shear flow, the fluid is known to show strong non linear behaviour with the existence of a stress plateau characteristic of the shear-banding transition \cite{Cap}. 
\subsection{\label{set}Setup}
\subsubsection{\label{cou}The Couette cell}
Our experiments are performed in a Perspex Couette device built in our lab and specially designed for the observations of the velocity gradient - vorticity ($\vec v,z$) plane (Fig. 1). The sample is placed in a gap $e$ of 1.13 mm between two concentric cylinders. The inner rotating cylinder has a radius $R_{1}$=13.33 mm and a height $h=40$ mm. It is of Mooney-Couette type with a cone-shaped lower part. The outer cylinder has a constant thickness of 2 mm and is surrounded by water which keeps the sample at constant temperature. The small tank, which allows the water circulation, is equiped with two glass windows, one for the radial incident beam and the other one for the direct observation of the gap. This configuration of the tank reduces the distortions of the image due to refraction effects on the successive cylindrical interfaces.\\
The cell is closed by a small plug which limits the destabilization processes of the free surface of the fluid at high strain rates. This allows to reach more easily the second stable branch without forming foam or bubbles, what often occurs with semi-dilute micellar samples \cite{Ber1}. We checked that the interfacial instability we observe between shear bands is not affected by the presence of the plug. A home-made solvent trap is also used to limit evaporation of the sample.
\subsubsection{\label{opt} The rheo-optical device}
Our transparent Couette cell is fitted to a stress-controlled rheometer (Physica MCR 500) also working in strain-controlled mode via a soft-controlled feedback loop, allowing to record the mechanical and optical responses simultaneously. A He-Ne laser ($\lambda=633$ nm) propagating along the velocity gradient axis is used as light source for the visualizations of the gap. By means of a plano-convex cylindrical lens ($f=4$mm), the slightly divergent beam is widened behind the focal point to form a laser light sheet. The thickness of the sheet (1.5 mm) is determined by the diameter of the incident beam into the flat surface of the lens. The intensity distribution in the light sheet being of Gaussian shape, the distance between the focal point and the input window of the tank is such as the high intensity region of the sheet covers all the height of the cell. Finally, a digital camera with resolution 720$\times$ 576 pixels records the scattered intensity at 90$^\circ$ at a rate of 25 frames per second, giving a view of gap in the plane ($\vec v,z$). The field of observation is centered at halfway of the cell, and varies from 0.5 to 1.7 cm according to the chosen magnification. We applied a numerical algorithm to each frame in order to detect the interface. The wavelength of the dominant mode is computed by direct Fourier transform. We also deduce the evolution with time of the amplitude of the interface profile.
\begin{figure*}[t]
\begin{center}
\includegraphics[scale=0.9]{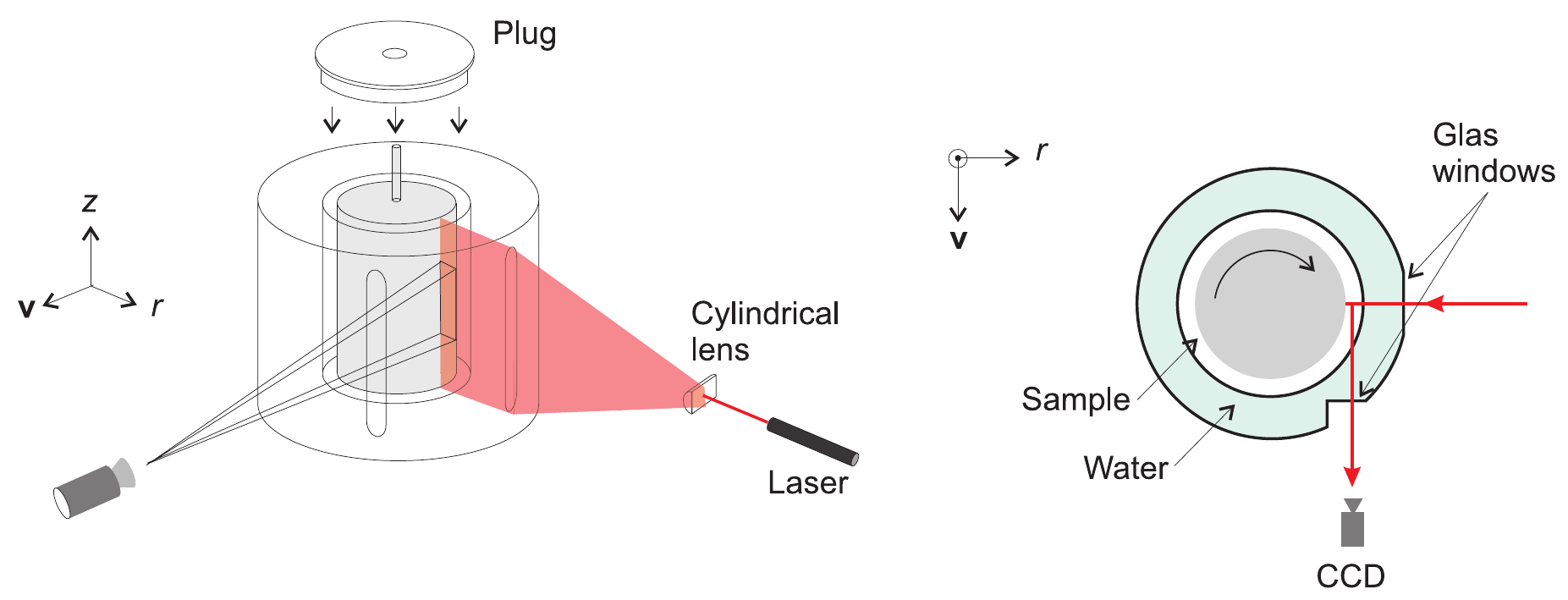}
\caption{\label{cou} (Color online)(a) Experimental setup for the observation of the gap in the plane ($\vec v,z$). (b) Top view of the Couette cell and of the measurement configuration.}
\end{center}
\end{figure*}
\section{\label{res}Results and discussion}
\subsection{\label{rheo} Mechanical rheology}
\subsubsection{\label{rheoa} Characterization of the stationary state}
\begin{figure}[b]
\begin{center}
\includegraphics[scale=0.9]{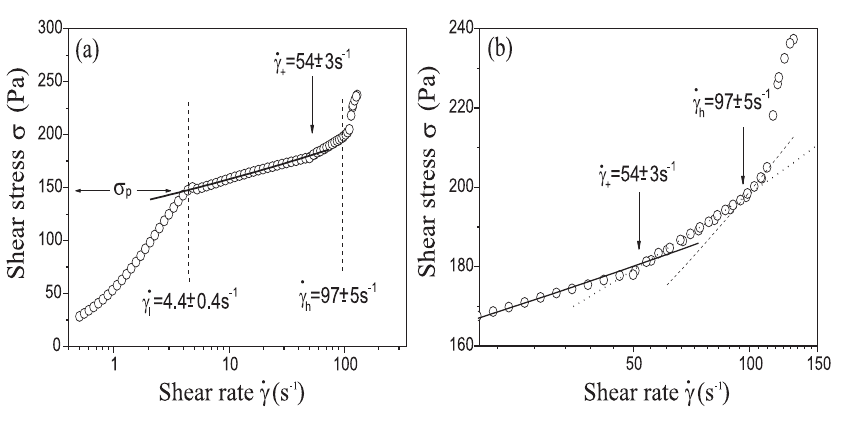}
\caption{\label{flowcurve} (a) Experimental steady-state flow curve measured under strain-controlled conditions. The sampling of the rate sweep is 180s per data point. (b) Enlargement of $\sigma(\dot\gamma)$ over $\dot\gamma=20-150s^{-1}$.}
\end{center}
\end{figure}

The steady-state shear stress as function of the shear rate obtained under strain-controlled conditions is shown in the semi-logarithmic plot in figure \ref{flowcurve}. The observed constitutive behaviour is in accordance with the non linear rheology of most of giant micelles systems \cite{Ber3} : the flow curve is made up of two increasing branches separated by a stress plateau at $\sigma_{p}=147\pm 0.5$ Pa extending from $\dot\gamma_{l}=4.4\pm 0.4s^{-1}$ to $\dot\gamma_{h}=97\pm 5s^{-1}$, characteristic of a shear-banding transition. A closer inspection of this graph deserves further comments. 
First, the stress plateau presents a significant positive slope. Assuming that the total stress tensor $\textbf{T}$ only depends on the radial coordinate \cite{Greco}, the momentum balance $\nabla \textbf{T}=0$ implies that, in the Couette geometry, the shear stress varies in the gap between the cylinders as:
\begin{eqnarray}
\sigma(r)=\dfrac{\Gamma}{2\pi hr^{2}}&=&\sigma(R_{1})\dfrac{R_{1}^{2}}{r^{2}}\\
&=&\dfrac{2R_{2}^{2}}{R_{1}^{2}+R_{2}^{2}} \sigma\dfrac{R_{1}^{2}}{r^{2}}
\end{eqnarray}
where $\Gamma$ is the torque measured on the axis of the inner cylinder, $h$ is the height of the cell, $\sigma(R_{1})$ the local stress at the rotor, and $\sigma$ the stress measured by the rheometer. As demonstrated in reference \cite{Sal2} with the hypothesis that the interface between bands is stable at a given local stress \cite{Lu, Ovi1, Ovi2}, this non homogeneity of the shear stress generates a stress increment $\Delta\sigma\approx 2e\sigma_{p}^{*}/R_{1}$  where $\sigma_{p}^{*}=2R_{2}^{2}\sigma_{p}/(R_{1}^{2}+R_{2}^{2})$ is the local stress at the moving wall associated with the onset of the plateau. We find $\Delta\sigma\approx 27$ Pa whereas the effective increase of the stress computed from the flow curve is about 50 Pa , indicating that the slope of the stress plateau is not fully explained by the curvature effects.\\
Second, the coexistence region is composed of two distinct parts, the small change of variation occuring around a shear rate $\dot\gamma_{+}=54\pm 3s^{-1}$ as illustrated in figure \ref{flowcurve}b. Each part can be described by a power law $\sigma=A\dot\gamma^{\alpha}$ with the following parameters: $A=130 Pa.s^{\alpha}$, $\alpha=0.08$ for $\dot\gamma_{l}\leq\dot\gamma<\dot\gamma_{+}$ and $A=100 Pa.s^{\alpha}$, $\alpha=0.14$ for $\dot\gamma_{+}\leq\dot\gamma\leq\dot\gamma_{h}$. This evolution is reproducible and we shall see later that the slight breaking off of the slope in the stress plateau at $\dot\gamma_{+}$ is concomittent with the appearance of a peculiar regime of spatio-temporal dynamics of the interface between bands, suggesting that the flow dynamics can play a role in the additionnal slope of the stress plateau. \\
Third, as mentioned in section \ref{cou}, the access to the second branch of the flow curve is often made delicate because of flow instabilities that generate bubbles and finally the expulsion of the sample from the gap of the cell. The plug on the top of our Couette device reduces the free surface of the fluid and shifts the instabilities thresholds towards higher shear rates. Then it becomes possible to provide a good approximation of the second critical shear rate $\dot\gamma_{h}$ and to describe the flow behaviour of the induced phase. We observe a deviation to the power law adjusting the second part of the stress plateau around $\dot\gamma=97\pm 5 s^{-1}$ and we assimilate this value with the shear rate $\dot\gamma_{h}$ corresponding to the upper limit of the stress plateau. This estimation is corroborated by the direct visualizations of the sample under shear (see section \ref{opt}), which show that all the gap is filled with the induced phase above $97 s^{-1}$. The value of the standard deviation is estimated from these optical observations and mainly takes into the account the fact that the precise shear rate for which the interface becomes perfectly flat and effectively touches the fixed wall is difficult to determine.
\begin{figure}[h]
\begin{center}
\includegraphics[scale=1.2]{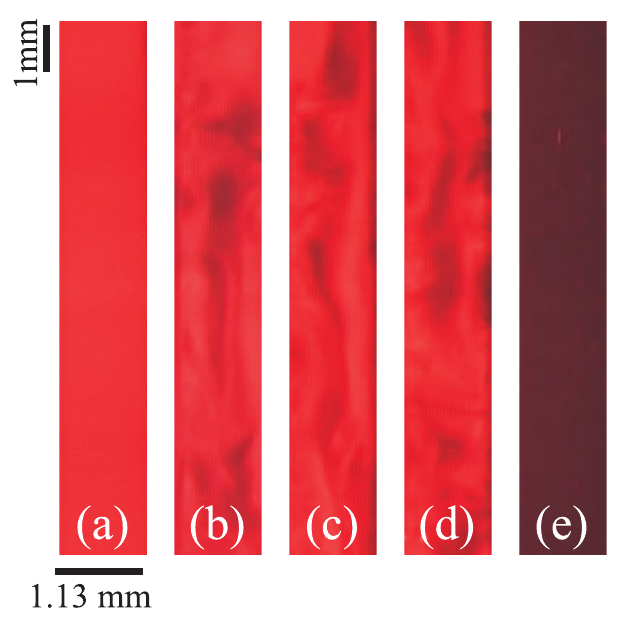}
\caption{\label{turb} (Color online) Snapshots of the plane $(r,z)$ of the Couette cell illuminated by a radial laser sheet : (a) The imposed shear rate is 105 $s^{-1}$ and lies in the shear-thinning region of the second branch. The induced phase completely  fills the gap and appears homogeneous. (b-d) Irregular patterns of the induced structure at $\dot\gamma=130s^{-1}$ in the shear-thickening regime along the second branch. (e) For comparison, a view of the gap at 2 $s^{-1}$, along the low viscosity branch.}
\end{center}
\end{figure}
From $\dot\gamma_{h}$ to 110 s$^{-1}$, the second branch of the flow curve follows a power law regime with $A=61 Pa.s^{\alpha}$ and $\alpha=0.26$, indicating that the induced phase is strongly shear-thinning, a feature already  highlighted by Salmon and coworkers on the classical CPCl/NaSal(6.3\%) system using both global rheology and local velocimetry measurements \cite{Sal1}. In this small range of shear rates, the induced phase appears homogeneous (see figure \ref{turb}.a). Above 110$s^{-1}$, the sample adopts a shear-thickening behaviour : in that case, the snapshots (fig.\ref{turb}.b-d) give a visual impression of strong spatial disorder in the gap of the cell, typical of random flow, this disorder nucleating systematically from the lower and upper boundaries of the inner rotating cylinder before spreading over all the height of the cell. This picture together with the increase of the torque as function of the shear rate is reminiscent of the state of elastic turbulence extensively studied by Groisman and Steinberg \cite{Groisman1} on dilute viscoelastic polymer solutions, the shear-thickening evolution being interpreted in terms of an increased flow resistance. Since such a phenomenon is driven by normal stresses effects, our observations tend to suggest that the induced structures are viscoelastic. Nonetheless, the window of shear rates that can be explored above 110 $s^{-1}$ is tiny since the flow is rapidly perturbed by bubbles, making a systematic study of the second branch difficult.
\subsubsection{\label{rheob} Time-dependent experiments}
\begin{figure}[h]
\begin{center}
\includegraphics[scale=1]{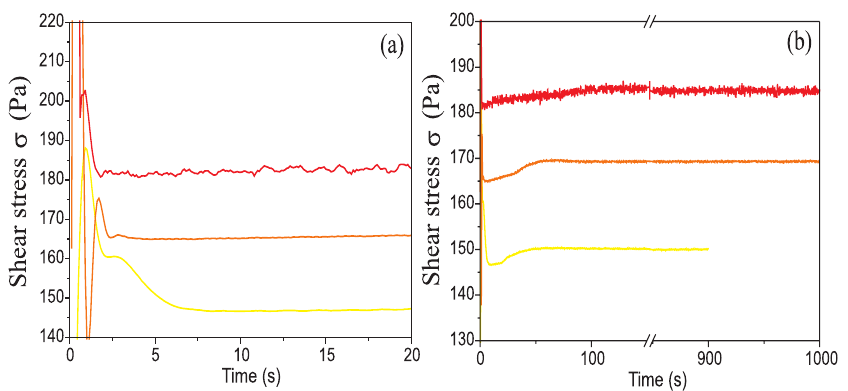}
\caption{\label{transient1} (Color online) Time-dependent response of the shear stress $\sigma(t)$ after the start-up of flow for several imposed shear rates along the stress plateau $\dot\gamma=7s^{-1}$ (yellow line),  $\dot\gamma=30s^{-1}$ (orange line),  $\dot\gamma=70s^{-1}$ (red line) (a) At short times. (b) On a longer time-scale to capture steady-state. }
\end{center}
\end{figure}

Figure \ref{transient1} displays the evolution of the shear stress as a function of time towards its steady state after a sudden start-up of flow for three different shear rates along the coexistence zone and on different time scales. The temporal stress response recorded with a sampling of 0.1s, is typical of a sample undergoing a transition of shear-banding type \cite{Ber2,Gra1,Ler1,Ler2}, with an overshoot followed by a relaxation to an apparently stable value in stretched exponential or damped oscillations depending on the applied shear rate (Fig.\ref{transient1}.a). The subsequent behaviour is a slow increase of the shear stress to its steady state value with a relative stress increment of about 2.5\% and can be related to the small undershoot already observed on other surfactant systems \cite{Ber2,Gra1,Ler1,Ler2} (Fig.3.b). \\

\begin{figure}[h]
\includegraphics[scale=1]{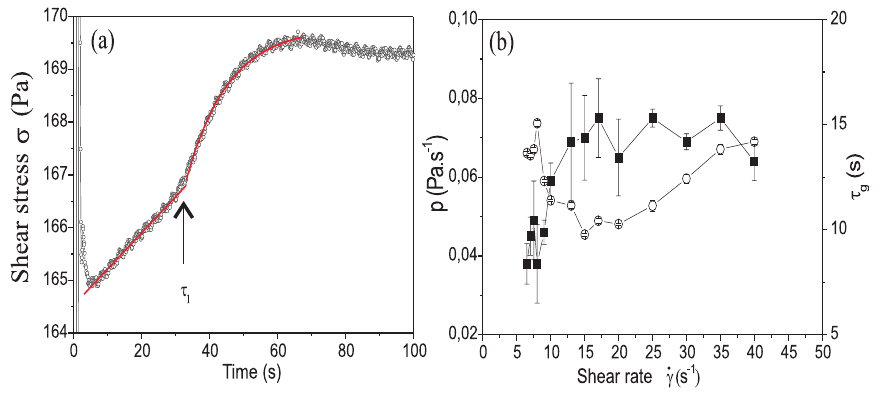}
\caption{\label{transient3} (Color online) a) View of the small undershoot in the stress response during the startup of flow at 30$s^{-1}$. For $t<\tau_{1}$, the stress increases linearly whereas above $\tau_1$, the variation can be fitted by a monoexponential growth. b) Slope $p$  ($\blacksquare$) of the linear law and characteristic time $\tau_g$ ($\circ$) of the exponential growth computed from the fitting procedure.}
\end{figure}

\begin{figure}[t]
\begin{center}
\includegraphics[scale=1.2]{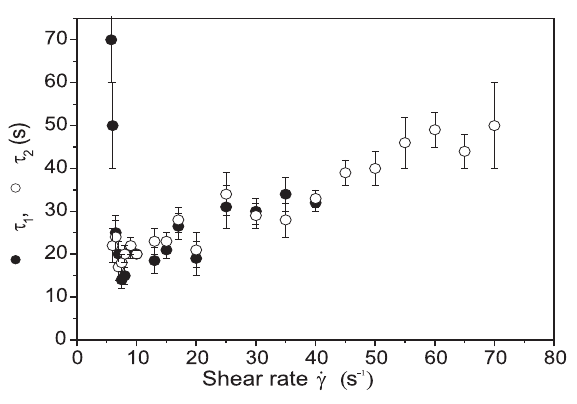}
\caption{\label{times} Comparison between the times $\tau_{1}$ (full symbols) extracted from the stress signal and the time $\tau_{2}$ (open symbols) characterizing the beginning of the destabilization of the interface. }
\end{center}
\end{figure}

A closer inspection of this part of the stress dynamics reveals different features depending on the applied shear rate. 
For most of the strain rates ranging from $\dot\gamma_{l}$ to 45s$^{-1}$, we observed generally two distincts variations as shown in figure \ref{transient3}.a at $\dot\gamma=30s^{-1}$ : the slow stress growth towards the steady state follows at first an affine law with a slope $p$ and then a monoexponential increase with a characteristic time $\tau_{g}$, the crossover between these two regimes occuring at a time named $\tau_{1}$. For some shear rates, the monoexponential growth is prolounged by a slight decrease towards steady-state (see fig.\ref{transient3}.a). The parameters $p$ and $\tau_{g}$ computed from the fitting procedure are gathered in figure \ref{transient3}.b. Except at low shear rates, the slope $p$ is approximately constant whereas the time $\tau_{g}$ is of the order of 10-15s. Let us mention that for some imposed shear rates, the kink at $\tau_{1}$ is not always so marked as in figure \ref{transient3}.a and in a few cases, the stress profile might also be fitted by a sigmo\"idal shape. 
Moreover, the time $\tau_{1}$ and the fitting parameters have been found to depend on the flow history : $\tau_{1}, p, \tau_{g}$ and the time characterizing the steady state increase with the time during which the sample stays at rest between two consecutive tests. The study of the effect of this lag time could maybe bring information on the relaxation of the induced structures and is left for a future work. In the present investigation, two consecutive tests have been systematically performed with a fixed lag time of the order of two minutes, namely much larger than the own time of the system and enough to ensure the reproducibility between the different tests.\\
Above 45s$^{-1}$, the stress growth becomes very slow and it is really difficult to extract a clear variation law. 
It is not possible to define $\tau_{1}$ either because the kink is buried in the shear stress fluctuations or because the variation laws of the slow stress growth towards the steady state differ from those below 45s$^{-1}$. In the next section, we shall try to identify the processes associated with these particular evolutions and we shall see that, whatever the averaged shear rate may be, the small undershoot in the shear stress appears as the mechanical signature of the interface instability between bands.\\
Figure \ref{times} shows that $\tau_{1}$ increases with the mean shear rate once it exceeds 10s$^{-1}$ which corresponds approximately to the spinodal point in the underlying flow curve \cite{Ber2,Gra1}. At low shear rates, we observe an upturn towards higher values of $\tau_{1}$ due to the existence of metastable states.\\
Finally we can note that the stress signal fluctuates around its steady-state value, the amplitude of the fluctuations growing with the imposed shear rate from 0.05\% at the beginning of the plateau to 0.3\% above 60$s^{-1}$, while the temporal fluctuations of the controlled shear rate never exceed 0.05\%. Such variations in the fluctuations of $\sigma(t)$ are probably related to the different regimes of spatio-temporal dynamics of the interface observed all along the stress plateau (see section \ref{spa}).
\subsection{\label{opt} Optical results}
\subsubsection{\label{cin} Observation of the banding structure}
\begin{figure*}[t]
\begin{center}
\includegraphics[scale=1]{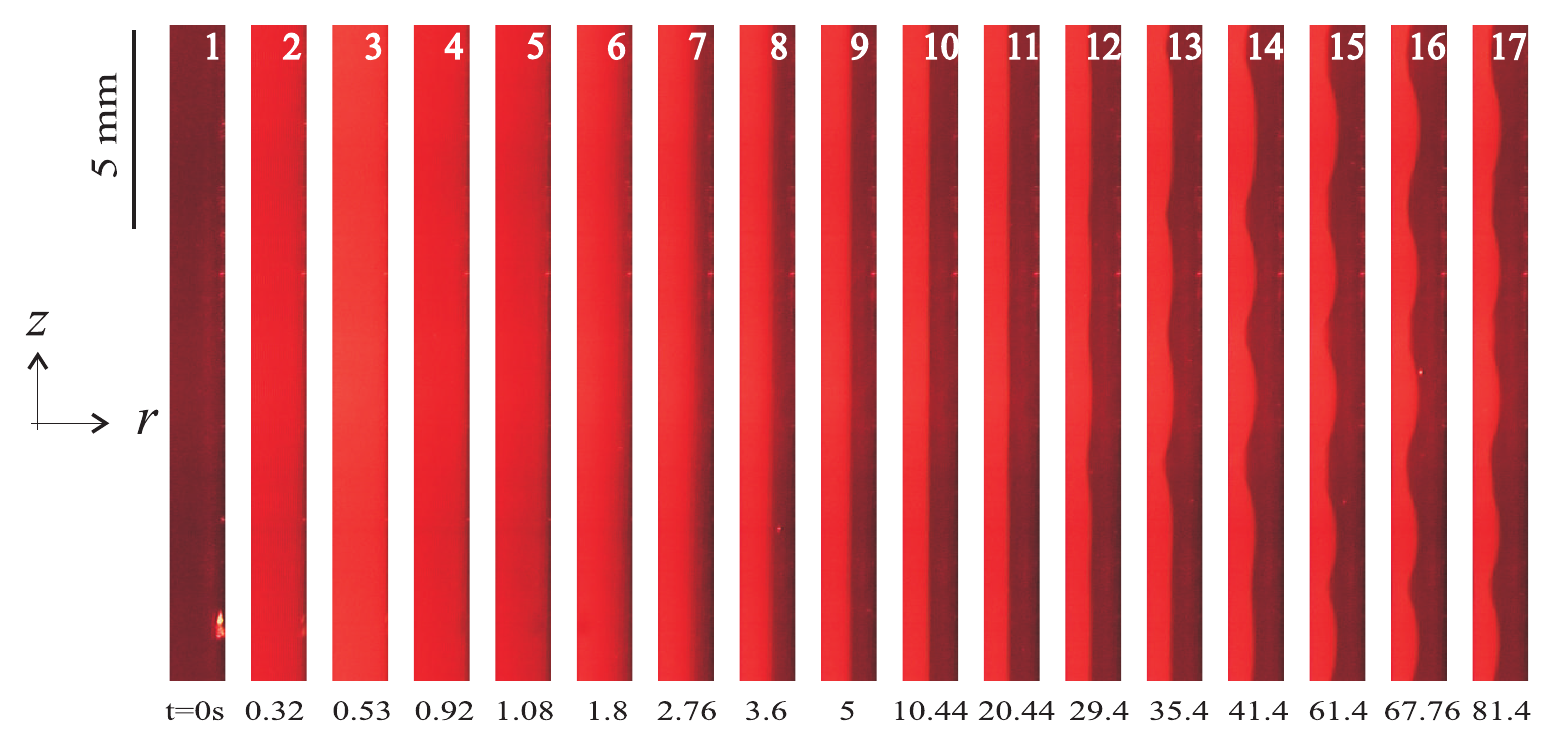}
\caption{\label{transient2}  (Color online) Photos of the gap of the Couette cell in the plane ($r,z$) after a startup experiment at 30$s^{-1}$. The left and right sides of each picture corresponds respectively to the inner and the outer cylinders. The height of the field of observation is centered at halfway of the measuring cell and limited to about 16 mm in order to keep a reasonable spatial resolution for the post-processing. At time $t=0$, the sample is at rest.} 
\end{center}
\end{figure*}
Figure \ref{transient2} illustrates the main stages of the kinetics of formation of both the banding state and the interfacial instability for a controlled shear rate of 30s$^{-1}$. Let us recall that the scattering signal is gathered simultaneously with the temporal stress evolution allowing thus a precise correlation of the structural and mechanical  responses. The sample is initially at rest and does not scatter the laser light (photo 1). At the onset of the simple shear flow, the entire gap becomes turbid, the maximum of scattered intensity being reached around 0.5s when the stress overshoot occurs (photo 3). Birefringence measurements have demonstrated that this non linear elastic response is associated with a strong orientation and stretching of the micellar threads with respect to the flow direction \cite{Ler2}. The observed turbidity results from the stretching of the micellar network that generates concentration fluctuations along the flow direction \cite{Hel} as suggested by small angle light scattering experiments under shear \cite{Ler4,Hu1}. According to Kadoma and coworkers,  a semi-dilute surfactant solution presents heterogeneties in micelles concentration at the scale of a few mesh sizes with regions of different densities of entanglements \cite{Kad}. The application of the flow field preferentially deforms the regions of low density of entanglements and produces the formation of clusters of mesoscopic size, first organized parallel to the flow direction. At this time, all the new phase is nucleated but not arranged into a macroscopic band.\\
The building of the banding structure starts with the relaxation of the stress overshoot : the turbidity first diminishes in a homogeneous way. This may be due either to the relaxation or to the breakup of the stretched micellar threads between the clusters. Then from 1.8s, one can observe the formation of a diffuse interface that begins to migrate from the fixed wall to its stationnary position in the gap (see photos 6 to 9).  The corresponding  behaviour in the shear stress response is the sigmo\"idal relaxation or the damped oscillations depending on the magnitude of the averaged shear rate. The migration is performed first rapidly with an approximate speed of 0.05 mm.s$^{-1}$ and then more slowly above $t$=5s. This process is accompagnied by a sharpening of the interface, the profile of which is flat on the total height of the Couette cell.\\
\begin{figure}[h]
\begin{center}
\includegraphics[scale=0.9]{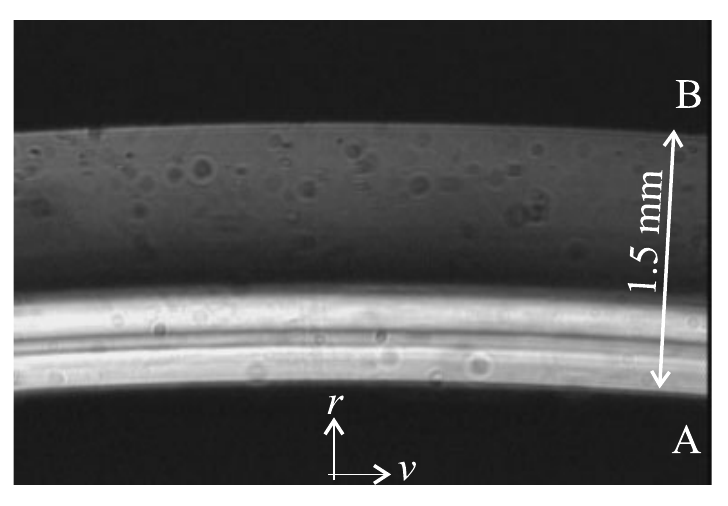}
\caption{\label{photo} Snapshot of the top of the Couette cell in the plane ($\vec v,r$) placed between crossed polarizer and analyzer and illuminated using an extended white light source. The letters A and B indicate the moving and the fixed cylinders respectively. The radius of the inner cylinder is 24.5 mm giving a gap of 1.5 mm and the height of the cell is 30 mm. The main axis of the polarizer  is orientated at approximately 20$^{\circ}$ with respect to the flow direction. The applied shear rate is 30s$^{-1}$.}
\end{center}
\end{figure}
Moreover, the band located near the fixed cylinder does not present any particular scattering properties as the sample at rest. However, this band is birefringent (see figure \ref{photo}), suggesting that it contains the initial entangled micellar network orientated by the local flow field. On the other hand, the induced band remains turbid, the scattered intensity level being uniform along the direction of the cylinder axis. The orientation and the stretching of the micron-sized domains created at the onset of flow could generate a form dichro\"ism related to the anisotropy of the scattered light. It has been shown recently that appearance of flow-induced dichro\"ism in wormlike micellar systems is  responsible for the turbidity \cite{Ful,Shu}. This form effect has to be studied in a careful way : it appears necessary to probe uniquely the induced band using small angle light scattering experiments under shear to determine quantitatively the characteristic size of the clusters and the way of which they interact. This shall help to establish the nature of the induced structure and is left for a future work. The induced band is also strongly birefringent as illustrated in figure \ref{photo}, reflecting a prononced orientation state of the micellar threads in the clusters. Let us note that form dichro\"ism is usually associated with form birefringence. However, the experiment between crossed polarizers does not allow to discriminate between form and intrinsic effects.\\
From $t\simeq$8s, the front adopts a sharp profile and continues to move at a speed varying from 0.007 to 2.10$^{-4}$ mm/s (see the inset in fig.\ref{position}) up to a time $\tau_{2}\simeq29$s where it seems to have almost reach its equilibrium position associated with the plateau value. During this slow displacement, the interface position evolves exponentially  (see figure \ref{position}) and the shear stress increases almost linearly as function of time (see figure \ref{transient3}.a). Let us mention that the period of the oscillations modulating the exponential evolution in figure \ref{position} is due to the imperfect rotation of the inner cylinder (the default of coaxiality inherent to the the apparatus is of the order of 20 $\mu m$). \\
\begin{figure}[h]
\begin{center}
\includegraphics[scale=1.2]{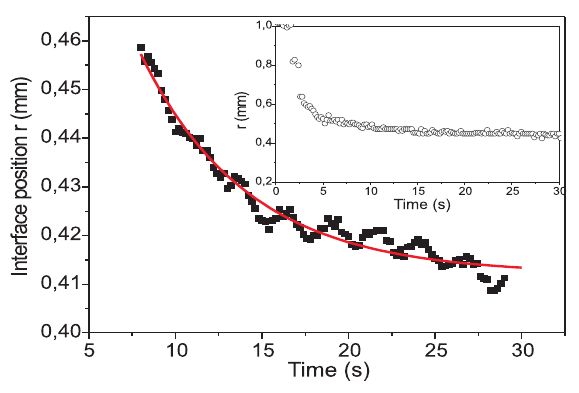}
\caption{\label{position} (Color online) Position of the interface as function of time during its slow migration towards its stationary position in the gap of the Couette cell. This evolution is well-described by a single exponential decay (red line). The inset illustrates the position of the interface from $t$=0 s until a time slightly upper than $\tau_{2}$.}
\end{center}
\end{figure}
According to the model proposed by Radulescu et al \cite{Ovi3}, the front propagation towards the final equilibrium position $r^{*}$ begins when the interface fully sharpens and is controlled by the stress diffusion coefficient $\D$ included in the original Johnson-Segalman model \cite{Olm1}. In this framework, the solution of the equation of motion of the interface gives:
\begin{equation}
r(t)-r^{*}=(r(0)-r^{*})e^{-t/\tau'}
\end{equation}
with
\begin{equation}
\tau'=\frac{\tau_{R}^{2}eKG_{0}\dot\gamma}{\sqrt{\D\tau_{R}}\eta_{l}\dot\gamma_{l}\delta\dot\gamma}
\end{equation}
where $\dfrac{KG_{0}\tau_{R}}{\eta_{l}\dot\gamma_{l}}\simeq 0.3$ \cite{Ovi3}, $\delta\dot\gamma$ is the width of the stress plateau and $\eta_{l}$ the viscosity of the sample just before the transition. 
In our case, the characteristic time $\tau'$ extracted from the interface displacement as function of time is equal to 6.2$\pm 0.3$ s and the diffusion coefficient $\D$ computed from this last equation is found to be $7.10^{-11}m^{2}.s^{-1}$, giving a correlation length $\varsigma=\sqrt{\D\tau_{R}}\simeq 4\mu$m. This value is larger by three orders of magnitude than the diffusion coefficient estimated on the same sample at 30$^{\circ}$C by following the stress relaxation after a small step between two values of the shear rate in the coexistence regime \cite{Ovi3}. The model predicted a stress relaxation on three well distinct time scales and supposed that the slowest relaxation stage was associated with the travel of the thin interface towards its stationary position. However this treatment ignored the phenomenon of destabilization of the interface which seems to be in fact the slowest process in the transient stress as we will see just below. Hence, the strong deviation between the values of $\D$ obtained here and in ref.\cite{Ovi3} could certainly be explained by the growth of the interface instability. Let us note that recently, stress diffusion coefficients of the order of  10$^{-12}-10^{-11} m^{2}.s^{-1}$  have been computed for CPCl/NaSal solutions at two different concentrations using superposition rheology together with ultrasonic velocimetry \cite{Bal}. Correlation lengths deduced from such values of $\D$ are also in agreement with recent measurements in straight microchannel on the same systems \cite{Mas}.\\
At $\tau_{2}$, we observe the first signs of destabilization of the interface along the  cylinder axis. Then from $\tau_{2}$ the instability grows with time and finally saturates around $t=60-70$ s where the interface shows undulations with a well-defined wavelength and a finite amplitude. The periodic pattern spreads over on all the height of the inner rotating wall as displays in figure \ref{edge}. The amplitude of the interface profile is minimum on the edges of the inner cylinder. \\
Let us note that the boundary conditions have been changed in order to test their effects on the existence of the instability. Our sample was sheared with a classical inner cylinder, with or without the plug, and in a partially filled cell and in each case, we observe the destabilization of the interface between bands. For the moment, we do not examine the way in which the undulations of the interface are quantitavely affected by a modification of the boundary conditions, of the gap thickness, the height and the curvature of the cell.\\
\begin{figure}
\begin{center}
\includegraphics[scale=0.9]{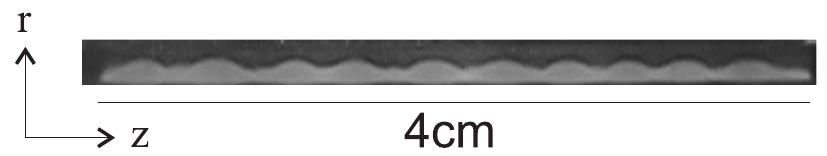}
\caption{\label{edge} Photograph of the plane $(r,z)$, the field of observation being extended on the total height of the Couette cell, namely 40mm.}
\end{center}
\end{figure}
In order to identify the part of the stress dynamics corresponding to the appearance and the development of the interface instability, we compared the time $\tau_{1}$ at which the kink between the linear and exponential regimes occurs during the slow growth of the shear stress towards steady state and the time $\tau_{2}$ at which the direct visualizations reveal the emergence of the instability (see figure \ref{times}). The correlation between both times is satisfying on the range $\dot\gamma\gtrsim\dot\gamma_{l}$ to 40 s$^{-1}$ taking into account the standard deviation on the measured values.\\ Just above the threshold $\dot\gamma_{l}$, the proportion of the induced band is extremely low and the resolution of our optical device does not allow for the detection of the interface profile while, from 45s$^{-1}$ and as mentioned in section \ref{rheob}, it is no more possible to define $\tau_{1}$ and we do not observe any particular changes during the slow stress growth around $t=\tau_{2}$. The existence of a complex dynamics of the interface above 45 s$^{-1}$ (see section \ref{spa}) could be responsible of these modifications. However, if the different regimes are not detectable in the stress signal at high shear rates, the instability still grows during the undershoot. \\
Hence, the undershoot in the stress curves as function of time (Fig.\ref{transient1}.b) appears as the mechanical signature of the interface instability along the cylinder axis. The evidence of such a characteristic in the transient stress profiles has been highlighted in other systems \cite {Gra1, Ber2} suggesting that this instability is not particular to our system. We shall show in the conclusion that the  CPCl/NaSal system, intensively studied by different groups during the last decade \cite{Ber2, Lopez1, Sal1}, also presents such an instability.\\
In the next section, we focus on the spatio-temporal dynamics of the interface all along the stress plateau.
\subsubsection{\label{spa} Phenomenology of the interface dynamics}
\begin{figure*}[t]
\begin{center}
\includegraphics[scale=0.6]{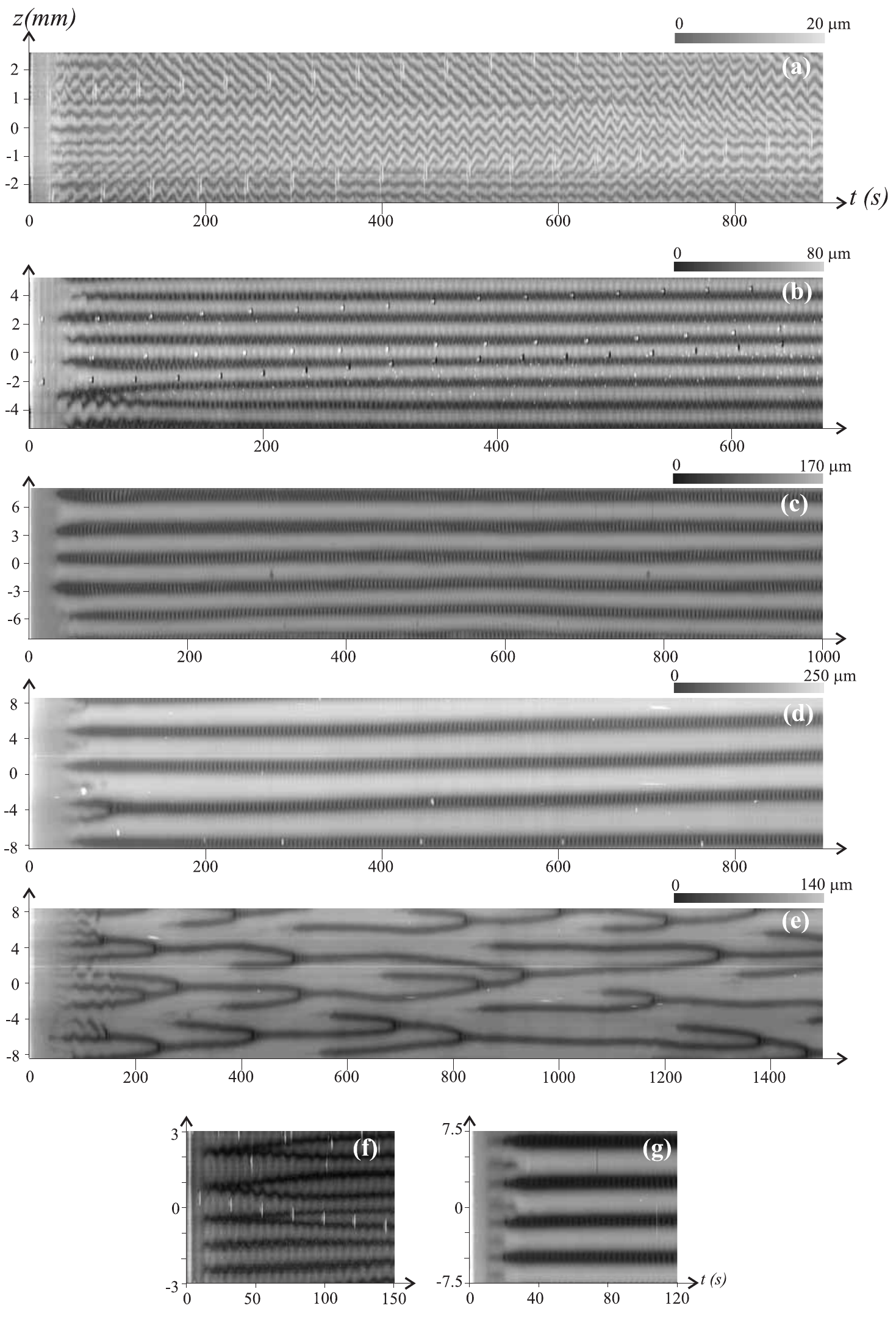}
\caption{\label{spatio28} Spatiotemporal evolution of the position of the interface in the gap of the Couette cell during a step shear rate from rest to (a) $\dot\gamma=6.5s^{-1}$, (b) $\dot\gamma=13s^{-1}$, (c) $\dot\gamma=30s^{-1}$, (d) $\dot\gamma=50s^{-1}$, (e) $\dot\gamma=70s^{-1}$. The position of the interface in the gap is given in grey scale, the origin being taken at the moving wall. The $z$ axis represents the spatial coordinate along the cylinder axis. Plots (f) and (g) correspond to applied shear rates of 13 and 50$s^{-1}$  at a temperature of 30$^{\circ}$.}
\end{center}
\end{figure*}

We follow the interface dynamics for a given step shear rate by collecting images of the gap in the plane ($r,z$) at a frame rate of 25 images per second. From each frame, we extract the interface profile and reconstruct the spatio-temporal diagram summarizing its evolution as function of the time and space coordinates. This  procedure is reproduced for various shear rates along the coexistence plateau and on a time scale allowing to capture both the transient regime and the asymptotic behaviour. Figure \ref{spatio28} gathers the different patterns that we have been able to identify. The grey levels materialize the position of the interface in the gap, black and white regions being associated respectively with the minima and maxima of the interface amplitude viewed from the moving cylinder. The $z$  coordinate corresponds to the direction of the cylinder axis, the origin being chosen at halfway of the cylinders.\\
Figure \ref{spatio28}.a displays the spatiotemporal sequence at $6.5s^{-1}$. After a transient of 24$\pm4$s including the construction, sharpening and migration of the interface, the pattern exhibits, at least at large scale, a well-defined wavelength $\lambda=0.5\pm 0.1$ mm, namely approximately half of the gap width. However, an inspection at smaller scale reveals a complex dynamics : globally, the pattern oscillates alternatively towards the top and the bottom of the Couette cell with a temporal period $T=17\pm1s$, which seems to be dissociated both of the period of rotation of the rotor (11.4s in that case) and of the time constant of the feed-back loop.
Moreover, waves propagating towards the bottom of the cell are clearly visible (see the top of the diagram). Let us note that propagative events towards the top of the cell simultaneously occur at other heights (not shown here), a feature already observed on the same system at a slightly different temperature \cite{Ler5}. At this shear rate, we computed an averaged phase velocity $v_{\phi}=18\mu m.s^{-1}$ with a standard deviation of 2$\mu m.s^{-1}$. This value is two orders of magnitude lower than the tangential velocity of the moving cylinder.\\  
This type of behaviour has been observed in the range of shear rates comprised between $\dot\gamma_{l}$ and 8s$^{-1}$. $v_{\phi}$ is found to increase with the macroscopic shear rate and reaches 34$\pm1\mu m.s^{-1}$ at the upper limit of domain (a) whereas $T$ diminishes but still in a decorrelated way with the rotation of the inner wall. Let us note however that, just above the threshold $\dot\gamma_{l}$, the proportion of the induced band is so reduced that the resolution of our CCD device forbids the quantitative description of the interface dynamics.\\
The subsequent behaviour of the interface is illustrated in figure \ref{spatio28}.b. In this range of shear rates  extending from 8 to 13s$^{-1}$, the dynamical evolution of the interface is mainly characterized by a decrease of the initially selected wavelength a few tens of second after the destabilization of the flat interface. This change of the wavelength is materialized by the development of new maxima (white zones) in the interface profile, as observed for example at 13s$^{-1}$ (Fig. \ref{spatio28}.b) around $t\simeq50$s. This effect is even more marked in figure \ref{spatio28}.f for the same solution but at a slightly different temperature : in this situation, we detect that the spatial frequency doubles. After this transient stage, the interface adopts a spatially stable profile with a wavelength $\lambda=1.48\pm0.03$mm.\\
Besides, the oscillation of the pattern towards the top and the bottom of the cell, typical of the previous dynamical regime, persists locally at short times before being finally damped. From the diagram at 13s$^{-1}$, the pseudo-period $T$ of the damped oscillations is estimated to 12.8$\pm$0.5s. For comparison, one rotation of the inner cylinder is performed  in 5.7s. \\
\begin{figure}
\begin{center}
\includegraphics[scale=1.2]{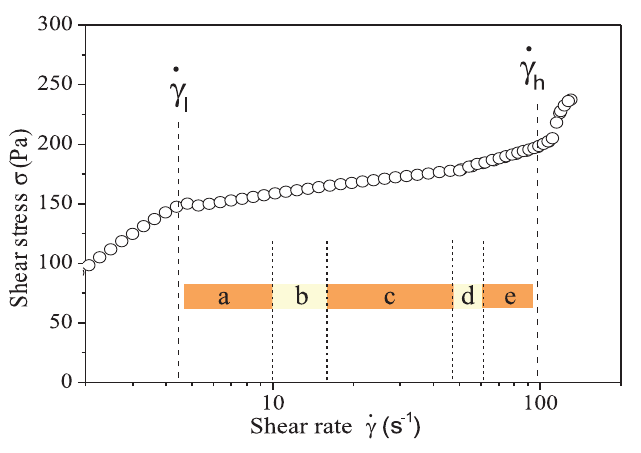}
\caption{\label{diag} (Color online) Illustration of the extension of each regime of interface dynamics along the coexistence plateau. }
\end{center}
\end{figure}
Figure \ref{spatio28}.c shows the dynamical behaviour encountered on a large part of the stress plateau from 13 to 45s$^{-1}$. In that case, the most amplified wavenumber in the initial stages of the instability growth is also the asymptotically dominant mode. The experiment at 30s$^{-1}$ has been carried out on a duration of one hour and we checked that there was no coarsening of the black and white regions : the interface keeps a spatially stable profile on very long times with a wavelength $\lambda=3.11\pm0.07$mm. A closer inspection of the diagram reveals that the amplitude of the interface profile is modulated in course of time : an oscillation of the minima (black zones) with a characteristic time of 5 s is clearly visible in figure \ref{spatio28}.c whereas the position of the maxima does not evolve at this particular frequency.  This phenomenon, detectable for shear rates ranging from 10 to 50s$^{-1}$, is also illustrated on the photographs in figure \ref{oscil} focusing on a minimum during a time scale of about 5 s. It is tempting to interpret this modulation of amplitude as a possible signature of instability with a wave vector orientated along the velocity. In the framework of the diffusive Johnson-Segalman model in the planar case, a linear stability analysis of a 1D shear-banding base flow recently predicted that the interface was unstable with respect to modes with wavevector along the velocity for a given range of interface sizes \cite{Fiel1,Wils1}. However, this assumption is difficult to defend in our case since the frequency of the modulation does not vary significantly with the macroscopic 
shear rate.
\begin{figure}
\begin{center}
\includegraphics[scale=0.5]{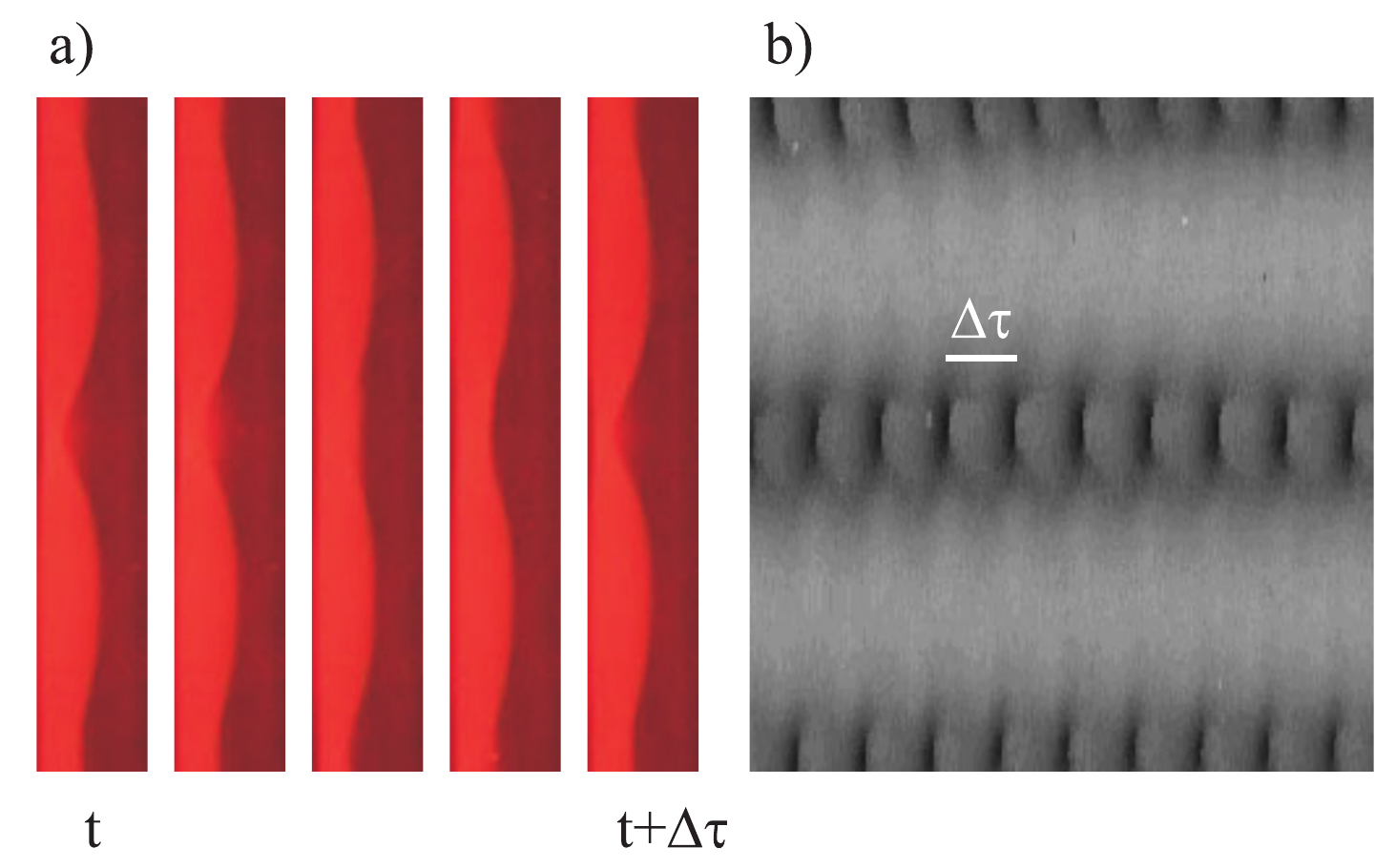}
\caption{\label{oscil} (Color online) a) Snapshots of one hollow of the interface profile at 30s$^{-1}$ in permanent regime on a period $\Delta\tau\simeq$ 5 s. b) Corresponding behaviour extracted from the spatio-temporal diagram at 30s$^{-1}$.}
\end{center}
\end{figure}
Besides, the coupling with the default of coaxiality intrinsic to the Couette cell when both frequencies are comparable, might affect the oscillation of the minima and makes the interpretation really complicated. Further experiments, for example based on observation of the interface in the $(\vec v,\vec\nabla v)$ plane have to be performed in order to understand the origin of this modulation. \\
The following regime (cf. Fig.\ref{spatio28}.d) is approximately confined between 45 and 55s$^{-1}$ and is characterized by a increase of the initially selected wavelength. In particular, at 50s$^{-1}$, the growth of the asymptotic mode is associated with a wavelength twice as large. This change of dominant mode occurs after about 50 to 60 s and beyond, the interface profile remains spatially stable with an asymptotic wavelength $\lambda=4.2\pm0.2$mm, namely about four times the gap width. Figure \ref{spatio28}.g provides an other illustration of the regime (d) at a temperature of 30$^{\circ}$C.\\
For $\dot\gamma\geq 55$s$^{-1}$, we observed a pronounced modification of the interface dynamics (cf. Fig.\ref{spatio28}.e). The emergence of this complex dynamics coincides with the change in the slope of the stress plateau at $\dot\gamma_{+}$ mentioned in section \ref{rheoa}.
The first signs of destabilization of the interface are detectable around 50 to 60 s, the length scale of the instability being of the order of 2 times the gap width at this moment. Let us note that this initial wavelength does not vary significantly with the imposed rate of strain. After a transient slightly lower than 200s, the amplitude of the interface profile saturates while the wavelength continuously evolves in course of time. In fact, two neightbouring hollows have a tendency to merge. When the length scale between two minima increases, several other hollows close to these minima nucleate and finally merge again with hollows of longer life time. This processes happen on all the duration of the experiment ($\simeq$25 min) : the system does not seem to tend toward a stationnary situation and the spatio-temporal diagram strongly suggests a chaotic dynamics.\\
When the applied shear rate is incremented along this region of the stress plateau,  the interface dynamics speeds up,  the life time of the minima being reduced and the number of union and nucleation events increasing notably. As in the first regime, it was not possible to describe the dynamics at the vicinity of the threshold $\dot\gamma_{h}$ because of the extremely low value of the amplitude of the interface profile.\\
\begin{figure}
\begin{center}
\includegraphics[scale=1]{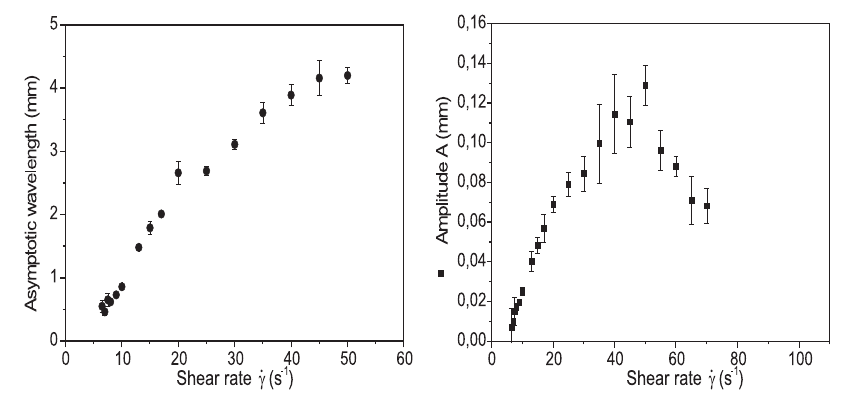}
\caption{\label{wave} (a) Asymptotic wavelength $\lambda$ ($\bullet$) versus $\dot\gamma$. The lack of experimental points above 50s$^{-1}$ is explained by the impossibility to define an asymptotic mode in regime (e). (b) Amplitude A of the interface profile ($\scriptscriptstyle\blacksquare$) as function of $\dot\gamma$.}
\end{center}
\end{figure}
Figure \ref{diag} summarizes the different regimes of interface dynamics along the coexistence plateau. On a great part of the stress plateau, the asymptotic pattern is stable in space (regions b, c and d), whereras the dynamics becomes complex when one approaches the thresholds (regions a and e), namely when the interface is close to the walls.\\
From the spatio-temporal diagrams, we extracted the wavelength and the amplitude of the dominant mode at long times and plotted their evolutions with the control parameter $\dot\gamma$ (cf. Fig.\ref{wave}). The asymptotic wavelength increases with $\dot\gamma$ and begins to saturate when one enters in regime (d). When $\dot\gamma$ rises above $\dot\gamma_{+}$, we cannot define an asymptotic wavelength. However, it is still possible to determine the order of magnitude of the "local" wavelength in this regime : the values are spaced out between 3 and 4 mm and globally decreases as the shear rate approaches $\dot\gamma_{h}$. In the range of strain rates where $\lambda$ is accessible, this latter does not seem to follow a simple scaling law with $\dot\gamma$. 
As for the amplitude of the interface, it follows a non-monotonic behaviour and presents a maximum for $\dot\gamma$ between 40 and 50 s$^{-1}$. Let us remark that the maximum value $A_{max}\simeq 120\mu m$ is reached when the two bands are in equal proportions in the gap (see figure \ref{wave} and \ref{wave1}). \\
In figure \ref{wave1}, we plotted the proportion of the induced band predicted by the lever rule $\alpha_{h}=(\dot\gamma-\dot\gamma_{l})/(\dot\gamma_{h}-\dot\gamma_{l})$. First, as already noted at another temperature \cite{Ler5}, the values computed from the latter equation with $\dot\gamma_{l}=4.4s^{-1}$ and $\dot\gamma_{h}=97s^{-1}$ (black line) are underestimated, even if we take into account the uncertainty on the critical shear rates (red dotted lines). Second, the experimental data do not seem to follow a linear behaviour with the applied shear rate contrary to the observations of Salmon et al on CPCl/NaSal \cite{Sal1}. Such deviation could perhaps be explained by the undulated profile of the interface along the $z$ direction. Besides, the existence of wall slip can not be excluded and could also justify that the measured proportion of the induced band is lower than the expected value. Let us finally  recall that we have considered that the turbid band was associated with the high shear rate band : for the moment, this hypothesis remains still to demonstrate by simultaneously measuring the local velocities and the optical properties. This type of experiments have been performed very recently on various CPCl/NaSal samples and they  indicate a strong correlation between shear, birefringence and turbidity banding. The authors also mention a breakdown of the lever rule \cite{Rau}.
\begin{figure}
\begin{center}
\includegraphics[scale=1.2]{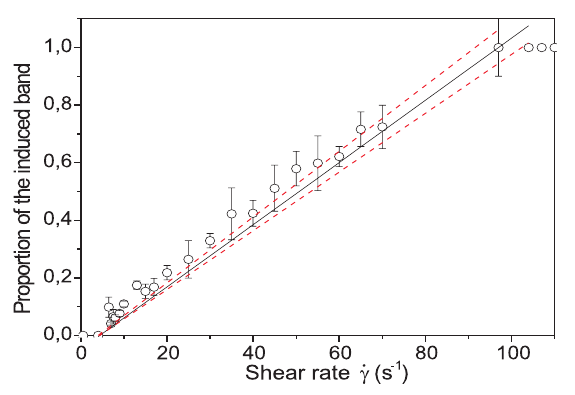}
\caption{\label{wave1} (Color online) Proportion $\alpha_{h}$ of the induced band ($\circ$) as function of $\dot\gamma$. The black  line represents the predictions of the lever rule with $\dot\gamma_{l}=4.4s^{-1}$ and  $\dot\gamma_{h}=97s^{-1}$. The dashed red lines show the extreme values of $\alpha_{h}$ taking into account the error bars on the critical shear rates.}
\end{center}
\end{figure}
\section{\label{conc} Conclusion}
\begin{figure}[h]
\begin{center}
\includegraphics[scale=0.5]{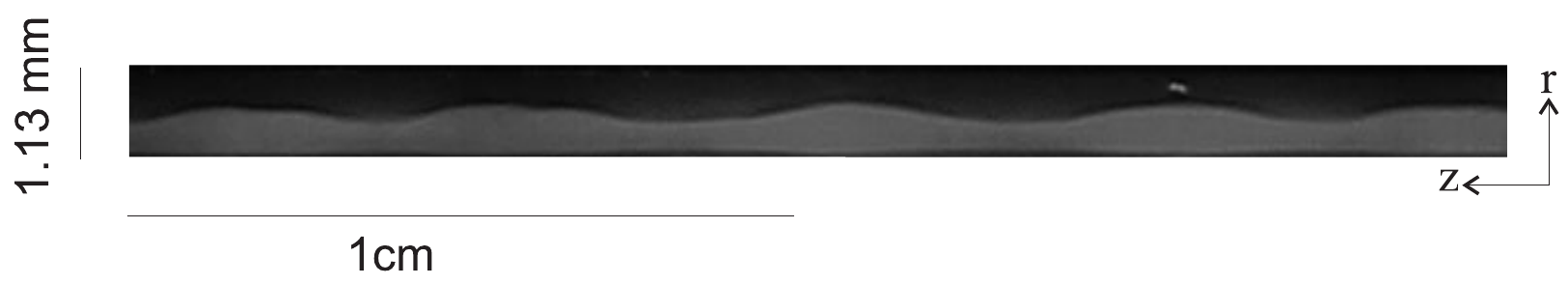}
\caption{\label{cpcl} Interface profile in a plane ($r,z$) for a sample of CPCl/NaSal (6.3\%) in 0.5 M NaCl brine at T=21.5 $^{\circ}$C. The mean shear rate is $\dot\gamma=9s^{-1}$ and the snapshot is taken after the steady banded state is achieved. }
\end{center}
\end{figure}

In this paper, we have highlighted a complex shear-banding scenario in which the steady banded state is characterized by an interface between bands undulating along the cylinder axis. Recent theoretical papers dealing with the stability of an initially 1D gradient banded base flow in the framework of the diffusive Johnson-Segalman model \cite{Fiel1,Wils1,Fiel2,Fiel3} have demonstrated that for sufficiently thin interfaces, the 1D state is unstable with respect to modes with wavevector along both the flow and the vorticity directions. Let us compare some of the predictions of the model in ref. \cite{Fiel3} with our results. First, the instability along the vorticity direction is predicted to grow much slower but in the asymptotic state, the  amplitudes of the oscillations in each of these directions are of the same order of magnitude \cite{Fiel3}. Taking into account the values of the amplitude gathered in figure \ref{wave}, the same type of effect along the velocity direction should be detectable. However, for the moment, we just detect a temporal modulation of the amplitude, the origin of which is not established and that does not seem to be compatible with undulations in the flow direction. To answer this question, we plan to modify our Couette device in order to visualize the interface profile in the plane of a laser sheet perpendicular to the cylinder axis.\\
Second, the comparison between the transient evolution of the shear stress and the kinetics of the instability allowed us to identify the slow stress growth towards steady state as the mechanical signature of the interface instability. This behaviour is quite  well captured by the model. This tends to point out that samples exhibiting such a feature in their stress time series are liable to undergo an interface instability. Figure \ref{cpcl} illustrates for example the steady state banding structure of a solution of CPCl/NaSal (6.3\%) in 0.5M NaCl brine at a mean shear rate of 9 s$^{-1}$. 
The interface between bands clearly undulates along the vertical axis,  the wavelength and the amplitude of the oscillation being respectively 4 mm and 120$\mu m$. We can also remark that the mechanical behaviour of the CTAB/KBr 
system is compatible with the existence of an interface instability \cite{Ler2}. In this system, a three-bands structure has been observed using the flow birefringence technique. In that case, the birefringence signal is averaged on the total height of the cell. Hence  the intermediate band in this particular banding structure could presumably result from the undulations of the interface along the vertical direction. This means that scattering or birefringence data resulting from an averaging along the vertical axis in Couette geometry could lead to misleading interpretation and should be examined with care. Let us mention that the slow growth of the mechanical stress also occurs in the shear-banding flow of onions \cite{Sal2} or  telechelic polymers \cite{Scho1}, making these complex fluids potential candidates for the interface instability. \\
Third, besides similar orders of magnitude for the wavelength and amplitude in the final state, at first sight, the evolution of the asymptotic wavelength with the average shear rate seems to be at least qualitatively reproduced by  the model (see the inset in figure 2 of ref. \cite{Fiel3}). Nevertheless, it would be necessary to compute it for smaller interface width to properly test the agreement on a much larger range of control parameter. \\
Fourth, one of the crucial point described by the model is the structure of the flow field in the plane ($r,z$). The scenario proposed by the author is as follows : the origin of the interface instability comes from the existence in the unperturbed flow of discontinuities across the interface in shear rate and in the normal stress parallel to the interface \cite{Fiel1,McLeish1,Hinch1}, the effects of these two driving terms being intertwined and not clearly understood yet. When the interface is subjected to a small perturbation, it becomes inclined, its normal $\textbf{n}$ being no longer radial. In order to keep the total velocity and the total traction ($S.\textbf{n}$, where $S$ denotes the unpertubed polymer stress and the stress perturbation) at the interface continous, a perturbation of the flow field must develop, mass conservation inducing recirculation which enhances the initial perturbation of the interface. In turn, Taylor-like vortices stacked along the vorticity direction form in the gap of the Couette cell, the size of the rolls scaling with the gap width. The centers of the rolls are localized at the vicinity of the interface, in the regions where this one is tilted. \\
\begin{figure}[h]
\begin{center}
\includegraphics[scale=1]{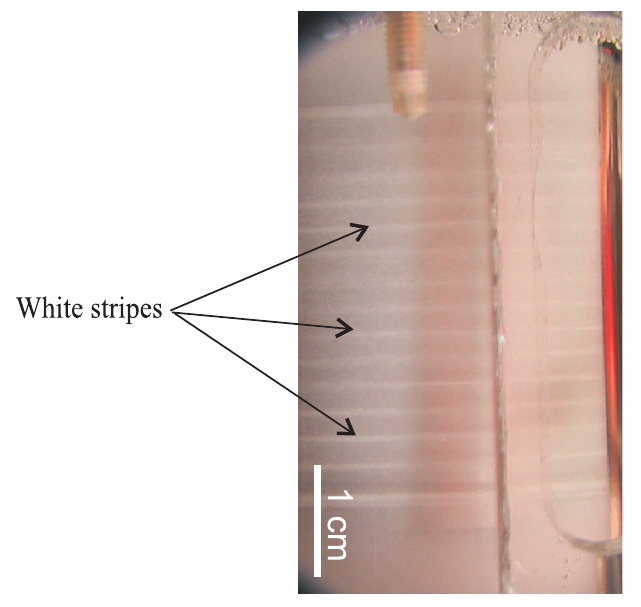}
\caption{\label{billes} External view of the Couette cell under ambient illumination at 20 min after the start-up of flow at $\dot\gamma=30s^{-1}$. Glass spheres of mean size 30 $\mu m$ are seeded at a concentration of 1\%, giving to the sample a milky aspect. The particles are organized into white stripes stacked along the vertical axis. }
\end{center}
\end{figure}
In order to test roughly the three-dimensional character of the flow field, we added small glass spheres (Sphericel, Potters industries) of mean radius 30$\mu m$ and density 1.1 at a concentration of 1\%. We checked that the mechanical properties of our sample were not affected by the particles. During a start-up of flow experiment at $\dot\gamma=30s^{-1}$, we observed that the glass beads tend to collapse into white stripes stacked along the vertical direction (see Fig. \ref{billes}), the wavelength of the pattern reaching approximately 2.5 mm when the steady-state is achieved. Of course this basic experiment does not allow to conclude about the structure of the flow field but it strongly suggests that recirculations occur in the sample. \\
Let us also mention that streaming velocity gradient along the vorticity direction induced by an interface has also been reported in molecular-dynamics simulations of the coexisting paranematic and nematic liquid-crystal phases under shear flow \cite{Germ1}.\\
Despite of the number of our observations captured by the model in \cite{Fiel3}, an alternative mechanism can be invoked to explain the vortex flow and the interface instability : the purely elastic instability in curved geometry at low Reynolds and Taylor numbers \cite{Lar1,Sha1}. This instability, observed in dilute viscoelastic polymer solutions \cite{Mull1} is due to the stretching of the molecules along the streamlines. This stretching along curved streamlines leads to negative normal stress difference $N_{1}=\sigma_{\theta\theta}-\sigma_{rr}$, producing a volume force (hoop stress), which acts in a direction opposite to the centrifugal effects. Above a critical  Weissenberg number, the hoop stress overcomes the centrifugal force and the azimuthal base flow becomes unstable with the development of Taylor-like vortices following a complex dynamics depending on the viscosity of the fluid and on the applied shear rate   \cite{Baum1}. We emphazised in section \ref{rheoa} that the induced phase is elastic. Hence, one can presumably assume that such an instability can potentially be triggered at least in this phase, generating vortices that destabilize the interface between bands. It should be noticed that a similar argument has been advanced as a possible origin of three-dimensional flow in the concentrated wormlike CTAB/D$_{2}$O system \cite{Man1}. Contrary to the model in ref \cite{Fiel3}, this mechanism needs curvature of the shearing cell. A relevant way to test the hypothesis of the elastic instability is to play with the curvature of the Couette cell and eventually to look at the system in a straight channel.\\ 
To summarise, we have presented a complete rheo-optical study of the dynamics of the shear-banding flow in a semi-dilute micellar  system. We have shown that our sample was not following the classical scenario reported in reference  \cite{Sal0}. The interface  between shear bands is found to be unstable with respect to wavevector along the vorticity direction, the mechanical signature of this instability being characterized by a small undershoot in the stress response. This behaviour is extremely robust and does not seem typical to our sample. 
Except in the vicinity of the thresholds where the evolution of the interface is not accessible, we have identified a complex spatio-temporal dynamics all along the coexistence plateau using the shear as controlled parameter : we observed propagative events at low shear rates, stable oscillating modes at intermediate  shear rates and finally chaotic patterns for the highest strain rates. Let us mention that this complex dynamics is very similar to that observed in the damped Kuramoto Sivashinski \cite{Chate}. This latter has been first introduced  as a model for the transition to spatio-temporal intermittency. In reference \cite{Ler5}, the equation for the shear banding dynamics has been guessed because of the small aspect ratio of the cell and because of the natural translational symmetry of the cylinder. We are now working in deriving this model equation from fluid mechanics principles, assuming a simple  strain rate/stress function similar to the flow curve measured in the present article \cite{Med}.\\
Moreover, the organization into stripes of small particles embedded in the solution suggests that the flow is three-dimensional with Taylor-like velocity rolls stacked along the vorticity axis. Further work based on particle image velocimetry experiments are considered to fully determine the velocity profiles in a radial plane \cite{Far1}.\\  
The nature of the induced band remains an open problem. The observation of random flow along the second branch indicates that the new structures are elastic while the turbidity points out the existence of a micrometric length scale in the system. Small angle light scattering experiments under shear performed selectively in the induced band could bring information on the characteristic size responsible for this turbidity. However it will be presumably inadequate to determine the organization of this "phase" at smaller scale. \\\\
\textbf{Acknowledgements}\\
The authors thank J.L Counord for the building of the optical device, J.P. Decruppe for fruitful discussions and the ANR JCJC-0020 for financial support.

%The references should start on their own page.

%\clearpage
%
%\begin{thebibliography}{99}
%
%\bibitem{marker}
%
%Author(s), {\it Journal title}, Year, {\bf Volume}(Issue number), First page number.
%
%% BibTeX users can use a style-file for PCCP, which can be found on the CTAN internet site of LaTeX: http://www.tex.ac.uk/tex-archive/biblio/bibtex/contrib/chem-journal/pccp.bst
%
%\end{thebibliography}

%Please compile a list of all figure captions on a separate page:

\end{document}